\documentclass[twocolumn,showpacs,preprintnumbers,
amsmath,amssymb,aps,prd,nofootinbib,superscriptaddress,
eqsecnum]{revtex4}
\usepackage{graphicx,color}

\makeatletter
\renewcommand{\p@subsection}{}
\makeatother

\begin{document}

\title{
Fluctuations and the Phase Transition
in a Chiral Model with Polyakov Loops~\footnote{
  Contribution to the proceedings of the International Workshop
  on Hadron Physics and Property of High Baryon Density Matter,
  November 22-25, Xian, China.
}
}

\author{C. Sasaki}
\affiliation{%
Gesellschaft f\"ur Schwerionenforschung, GSI,  D-64291 Darmstadt,
Germany}
\author{B. Friman}
\affiliation{%
Gesellschaft f\"ur Schwerionenforschung, GSI,  D-64291 Darmstadt,
Germany}
\author{K. Redlich}
\affiliation{%
Institute of Theoretical Physics, University of Wroclaw, PL--50204
Wroc\l aw, Poland}

\date{\today}

\begin{abstract}
We explore the NJL model with Polyakov loops
for a system of three colors and two flavors
within the mean-field approximation,
where both chiral symmetry and confinement
are taken into account.
We focus on the phase structure of the model and study
the chiral and Polyakov loop susceptibilities.
\end{abstract}

\maketitle


\section{Introduction}
\label{sec:int}

Low-energy phenomena of QCD have been studied in various
effective models based on chiral symmetry.
However, the relation of spontaneous chiral symmetry breaking 
and confinement remains an open issue.
Recently, color degrees of freedom were introduced in the 
Nambu--Jona-Lasinio (NJL) Lagrangian~\cite{njl,review}
through an effective gluon potential expressed in terms
of Polyakov loops (PNJL model)~\cite{pnjl1,pnjl2}.
The basic ingredients of the model are constituent quarks and
the Polyakov loop, which is an order parameter of the $Z(3)$ 
symmetry of QCD in the heavy quark limit.
The model has a non-vanishing coupling of the constituent quarks
to the Polyakov loop and mimics confinement
in the sense that only three-quark states contribute to the 
thermodynamics in the low-temperature phase.
Hence, the PNJL model yields a better description of QCD 
thermodynamics near the phase transition than the NJL model.
Furthermore, due to the symmetries of the Lagrangian, the model 
belongs to the same universality class as that expected for QCD. 
Thus, the model can be considered as a testing ground for the 
critical phenomena related to the breaking of the global $Z(3)$ 
and chiral symmetries.

It has been shown that the PNJL model, formulated at finite 
temperature and finite quark chemical potential, 
well reproduces some of the thermodynamical observables 
computed within lattice gauge theory (LGT). 
The properties of the equation of state~\cite{pnjl2}, 
the in-medium  modification of meson masses~\cite{pnjl:meson} 
as well as the validity and applicability of the
Taylor expansion in quark chemical potential used in LGT were
recently addressed within the PNJL model
\cite{sus:pnjl,pnjl:iso,RRW:phase,RRW:qsus}. 
In Ref.~\cite{pnjl:isov} the
model was extended to a system with finite isospin chemical
potential and pion condensation was studied.

Enhanced fluctuations are characteristic for phase transitions.
Thus, the exploration of fluctuations is a promising tool for
probing the phase structure of a system. The phase boundaries can
be identified by the response of the fluctuations to changes in the
thermodynamic parameters.
In this article we describe the susceptibilities of the order
parameters and their properties in the PNJL model following
Ref.~\cite{SFR:PNJL}.


\section{Nambu--Jona-Lasinio model with Polyakov loops}
\label{sec:model}

An extension of the NJL Lagrangian by coupling the quarks to 
a uniform temporal background gauge field, 
which manifests itself entirely in the Polyakov loop,
has been proposed to account for interactions with
the color gauge field in effective chiral models~
\cite{pnjl1,pnjl2}. 
The PNJL Lagrangian for three colors ($N_c = 3$) and two 
flavors ($N_f = 2$) with non-local four-fermion
interactions is given by
\begin{eqnarray}
{\mathcal L}
&=& \bar{\psi} (i\gamma^\mu D_\mu - \hat{m})\psi
{}+ \bar{\psi}\hat{\mu}\gamma_0\psi
{}- {\mathcal U}(\Phi[A],\bar{\Phi}[A];T)
\nonumber\\
&&
{}+ \frac{G_S}{2}\left[\left(\bar{q}(x)q(x)\right)^2
{}+ \left(\bar{q}(x) i\gamma_5\vec{\tau}q(x)\right)^2 \right]\,,
\label{lag}
\end{eqnarray}
where $\hat{m} = \mbox{diag}(m_u, m_d)$ is the current quark mass,
$\hat{\mu} = \mbox{diag}(\mu_u,\mu_d)$ is the quark chemical 
potential  and   $\vec{\tau}$ are the Pauli matrices. 
We assume isospin symmetry and take $m_u = m_d \equiv m_0$ and 
$\mu_u = \mu_d \equiv \mu$.

Here we have introduced non-local interactions which are
controlled by a form factor in order to deal with the ultraviolet 
singularities that appear in the loop integrations. 
In coordinate space, the form factor $F(x)$ for the non-local 
current-current interaction reads:
\begin{equation}
q(x) = \int d^4y F(x-y)\psi(y)\,.
\end{equation}
A possible choice for the regulator is in momentum space given
by~\cite{nonlocal}:
\begin{equation}
f^2(p) = \frac{1}{1 + (p/\Lambda)^{2\alpha}}\,,
\end{equation}
where $f(p)$ is the Fourier transform of the form factor $F(x)$ and
$p$ is the three--momentum.
The NJL sector is controlled by four parameters: the strength of
four-fermion interaction $G_S$, the current quark mass $m_0$ and 
the constants $\alpha$ and $\Lambda$, which characterize the range 
of the non-locality. 
These parameters are determined in vacuum, for a given $\alpha$, 
by requiring that the experimental values of the pion parameters
and the quark condensate are reproduced.

The interaction between the effective gluon field and the quarks
in the PNJL Lagrangian is implemented (\ref{lag}) by means of a
covariant derivative
\begin{equation}
D_\mu = \partial_\mu - iA_\mu\,, \quad \quad A_\mu =
\delta_{\mu 0}A^0\,,
\end{equation}
where we introduce the standard notation $A_\mu = g A_\mu^a
\frac{\lambda^a}{2}$. Here $g$ is the color SU(3)
gauge coupling constant and $\lambda^a$ are the Gell-Mann matrices.

The effective potential ${\mathcal U}$ of the gluon field in
(\ref{lag}) is expressed in terms of the traced Polyakov loop
$\Phi$  and its conjugate  $\bar{\Phi}$
\begin{equation}
\Phi = \frac{1}{N_c}\mbox{Tr}_c L\,, \qquad 
\bar{\Phi} = \frac{1}{N_c}\mbox{Tr}_c L^\dagger\,, 
\label{phi}
\end{equation}
where $L$ is a matrix in color space related to the gauge field by
\begin{equation}
L(\vec{x}) = {\mathcal P}\exp\left[i\int_0^\beta d\tau
A_4(\vec{x},\tau)\right]\,,
\end{equation}
with ${\mathcal P}$ being the path (Euclidean time) ordering, 
and $\beta = 1/T$ with $A_4 = iA_0$.
In the heavy quark mass limit,  QCD has the Z(3) center symmetry
which is spontaneously broken  in the high-temperature phase. The
thermal expectation value of the Polyakov loop $\langle \Phi
\rangle$ acts as an order parameter of the Z(3) symmetry.
Consequently, $\langle\Phi\rangle = 0$ at low temperatures in the
confined phase and $\langle \Phi \rangle \neq 0$ at high
temperatures corresponding to the deconfined phase.

The effective potential $\mathcal { U}(\Phi,\bar{\Phi})$ of the
gluon field is expressed in terms of the Polyakov loops so as to
preserve the $Z(3)$ symmetry of the pure gauge theory~\cite{dumitru}. 
We adopt an   effective potential of the
following form~\cite{pnjl2}:
\begin{equation}
\frac{{\mathcal U}(\Phi,\bar{\Phi};T)}{T^4}
= - \frac{b_2(T)}{2}\bar{\Phi}\Phi
{}- \frac{b_3}{6}(\Phi^3 + \bar{\Phi}^3)
{}+ \frac{b_4}{4}(\bar{\Phi}\Phi)^2\,,
\label{eff_potential}
\end{equation}
with
\begin{equation}
b_2(T) = a_0 + a_1\left(\frac{T_0}{T}\right)
{}+ a_2\left(\frac{T_0}{T}\right)^2
{}+ a_3\left(\frac{T_0}{T}\right)^3\,.
\end{equation}
The coefficients $T_0$, $a_i$ and $b_i$ are fixed by requiring that
the equation of state obtained  in pure gauge theory on the lattice
is reproduced. In particular, at $T_0 = 270$ MeV the model
reproduces the first order deconfinement phase transition of the
pure gauge theory.


\section{Susceptibilities and the phase structure}
\label{sec:sus}

In the PNJL model the constituent quarks and the Polyakov loops are
effective fields related with the order parameters for the chiral
and $Z(3)$ symmetry breaking. In LGT the susceptibilities
associated with these fields show clear signals of the phase
transitions~\cite{lattice}.

\begin{figure}
\begin{center}
\includegraphics[width=8cm]{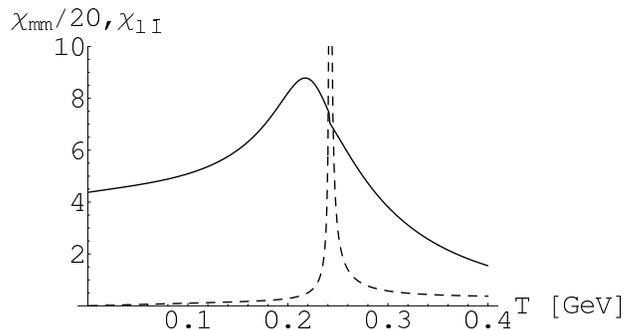}
\caption{
The chiral $\chi_{mm}$ (dashed-line) and the Polyakov loop
$\chi_{l\bar{l}}$ (solid-line) susceptibilities in the chiral limit
as functions of temperature $T$ for $\mu = 0$.
}
\label{fig:sus}
\end{center}
\end{figure}

In Fig.~\ref{fig:sus} we
show the chiral $\chi_{mm}$ and Polyakov loop 
$\chi_{l\bar{l}} = \langle \bar{\Phi}\Phi \rangle - 
\langle \bar{\Phi} \rangle \langle \Phi \rangle$ susceptibilities
computed at $\mu= 0$ in the PNJL model in the chiral limit 
within the mean field approximation.
The chiral susceptibility exhibits a very narrow divergent peak at
the chiral critical temperature $T_{\rm ch}$, 
while the peak of $\chi_{l\bar{l}}$ is much broader and the
susceptibility remains finite at all temperatures. This is due to
the explicit breaking of the $Z(3)$ symmetry by the presence of
quark fields in the PNJL Lagrangian. Nevertheless, $\chi_{l\bar l}$
still exhibits a peak structure that can be considered as a signal
for the deconfinement transition in this model.
One finds the interference of $\chi_{l\bar l}$ with $\chi_{mm}$ 
in Fig.~\ref{fig:sus}: 
At the chiral transition, $T=T_{\rm ch}$, there is a narrow structure
in $\chi_{l\bar l}$. We stress that this feature is not related
with the deconfinement transition, but expresses
the coupling of quarks to the Polyakov loops.
Thus, for the parameters used in the
model, the deconfinement transition, signaled by the broad bump in
$\chi_{l\bar l}$, sets in earlier than the chiral transition at
vanishing net quark density.

\begin{figure}
\begin{center}
\includegraphics[width=8cm]{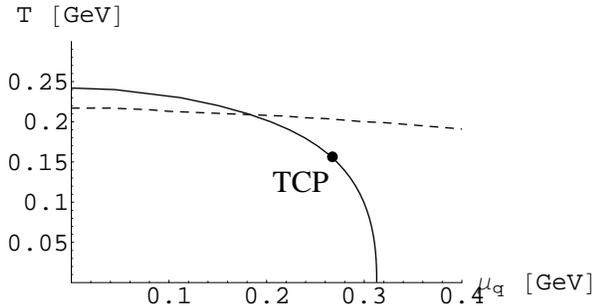}
\caption{
The phase diagram of the PNJL model in the chiral limit. The
solid (dashed) line denotes the chiral (deconfinement) phase
transition respectively. The TCP (bold-point) is located at
$(T_c=157 ,\mu_c=266)$ MeV.
}
\label{fig:phase}
\end{center}
\end{figure}

The peak positions of the $\chi_{mm}$ and $\chi_{l\bar{l}}$
susceptibilities determine the phase boundaries in the
$(T,\mu)$--plane. 
At finite chemical potential, there is a shift of the chiral phase
transition to lower temperatures.
In Fig.~\ref{fig:phase} we show the resulting phase diagram for the
PNJL model. The boundary lines of deconfinement and chiral symmetry
restoration do not coincide. There is only one common point in the
phase diagram where the two transitions appear simultaneously.
Recent LGT results both at vanishing and at finite quark chemical
potential show that deconfinement and chiral symmetry restoration
appear in QCD along the same critical line~\cite{lattice}. 
In general it is possible to choose the PNJL model parameters 
such that the both critical temperatures coincide at $\mu=0$.

From Fig.~\ref{fig:phase} one finds that
the slope of $T_{\rm dec}$ as a function of $\mu$ is almost flat,
indicating that at low temperature the chiral phase transition
should appear much earlier than deconfinement.
However, there are
general arguments, that the deconfinement transition should precede
restoration of chiral symmetry (see e.g.~\cite{Shuryak, Pisarski}).
In view of this, it seems unlikely that at $T\simeq 0$ the chiral
symmetry sets in at the lower baryon density than deconfinement.
In the PNJL model, the parameters of the
effective gluon potential were 
fixed by fitting quenched LGT calculations. 
Consequently, the parameters are taken as independent on $\mu$. 
However, it is conceivable  that the effect of dynamical quarks 
can modify the coefficients of this potential, thus resulting in 
$\mu$--dependence of the parameters.
Consequently, the slope of $T_{\rm dec}$ as a function of $\mu$
could be steeper~\footnote{
 Such a modification was explored in
 Ref.~\cite{mu-dep} where explicit $\mu$- and $N_f$- dependence of
 $T_0$ is extracted from the running coupling constant $\alpha_s$,
 using the argument based on the renormalization group.}.
Consequently, the effective Polyakov loop potential
(\ref{eff_potential}) should, with $\mu$-independent coefficients,
be considered as a good approximation only for $\mu/T<1$.

While the susceptibilities $\chi_{mm}$ and $\chi_{l\bar{l}}$ 
exhibit expected behaviors associated with the phase transitions,
other Polyakov loop susceptibilities,
\begin{eqnarray}
\chi_{ll}
&=&
\langle \Phi^2 \rangle - \langle \Phi \rangle^2\,,
\nonumber\\
\chi_{\bar{l}\bar{l}}
&=&
\langle \bar{\Phi}^2 \rangle - \langle \bar{\Phi} \rangle^2\,,
\end{eqnarray}
are negative in a broad temperature range above $T_{\rm ch}$
\cite{SFR:PNJL}. 
This is in disagreement with recent lattice results, where 
$\chi_{{l} {l}}$ is always positive in the presence of dynamical 
quarks~\cite{lattice}. 
A possible origin of this behavior could be the approximation to 
the effective Polyakov loop potential used in the 
Eq.~(\ref{eff_potential}).

\begin{figure}
\begin{center}
\includegraphics[width=8cm]{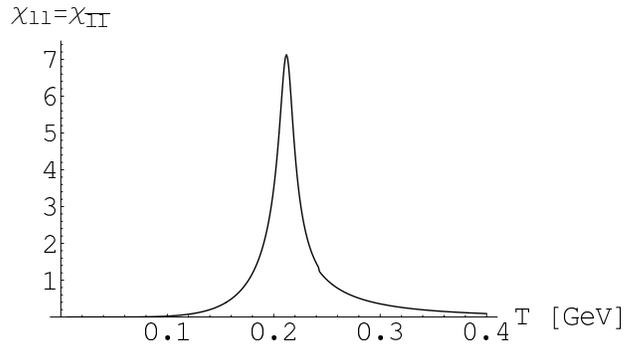}
\caption{\label{fig:susll_imp} 
The $\chi_{ll}=\chi_{\bar{l}\bar{l}}$ susceptibility
in the chiral limit as a function of temperature $T$ for $\mu = 0$.
The effective Polyakov loop potential (\ref{eff_imp}) was used.}
\end{center}
\end{figure}

Recently an improved effective potential with temperature-%
dependent coefficients has been suggested~\cite{RRW:phase}
\begin{eqnarray}
&&
\frac{{\cal U}(\Phi,\bar{\Phi};T)}{T^4}
= - \frac{a(T)}{2} \bar{\Phi}\Phi
\nonumber\\
&&
{}+ b(T) \ln \left[ 1 - 6\bar{\Phi}\Phi
{}+ 4\left( \Phi^3 + \bar{\Phi}^3 \right) - 3\left( \bar{\Phi}\Phi \right)^2 \right]\,,
\label{eff_imp}
\end{eqnarray}
where
\begin{equation}
a(T) = a_0 + a_1 \left( \frac{T_0}{T} \right) 
{}+ a_2 \left( \frac{T_0}{T} \right)^2\,,
\quad
b(T) = b_3  \left( \frac{T_0}{T} \right)^3\,,
\end{equation}
The polynomial in $\Phi$ and $\bar{\Phi}$,  used in
(\ref{eff_potential}), is replaced by a logarithmic term, which
accounts for the Haar measure in the group integral. 
The parameters in (\ref{eff_imp}) were
fixed to reproduce the lattice results for pure gauge QCD
thermodynamics and for the behavior of the Polyakov loop. 
In Fig.~\ref{fig:susll_imp} we show the  $\chi_{ll}$ susceptibility
calculated with the potential of Eq. (\ref{eff_imp}).
The improved potential indeed yields positive values
for all the Polyakov loop susceptibilities . 
We note that the phase diagram calculated with the improved 
potential is similar to that obtained with the
previous choice of the Polyakov loop interactions,
shown in Fig.~\ref{fig:phase}.


\setcounter{equation}{0}
\section{Summary and discussions}
\label{sec:sum}

We introduced susceptibilities related with the three different
order parameters in the PNJL model, and analyzed their properties and
their behavior near the phase transitions. 
In particular, for the quark-antiquark and chiral density-density 
correlations we have discussed the interplay between the restoration 
of chiral symmetry and deconfinement. 
We observed that a coincidence of the deconfinement and chiral 
symmetry restoration is accidental.

We found that, within the mean field approximation and with the
polynomial form of an effective gluon potential the correlations of
the Polyakov loops in the quark--quark channel show an unphysical
behavior, being negative in a broad parameter range. This behavior
was traced back to the parameterization  of the Polyakov loop
potential. We argue that the $Z(N)$-invariance of this potential
and the fit to lattice thermodynamics in the pure gluon sector is
not sufficient to provide correct description of the Polyakov loop
fluctuations.
Actually it was pointed out~\cite{RRW:phase} that the polynomial
form used in this work does not possess the complete group structure
of color SU(3) symmetry.
The improved potential of Ref.~\cite{RRW:phase}
yields a positive, i.e. physical, $\chi_{ll}$ susceptibility, 
in qualitative agreement with the LGT results.


\section*{Acknowledgments}

The work of B.F. and C.S. was supported in part
by the Virtual Institute of the Helmholtz Association under 
the grant No. VH-VI-041. 
K.R. acknowledges partial support of the Gesellschaft f\"ur 
Schwerionenforschung (GSI) and 
Committee for Scientific Research (KBN).


\end{document}